\batchmode
\makeatletter
\def\input@path{{/Users/ananyabalakrishna/Documents/Disemmination/Writing/Arxived/3_Periodic_patterns/Lyx/}}
\makeatother
\documentclass[10pt,twocolumn]{article}
\usepackage[latin9]{inputenc}
\usepackage[letterpaper]{geometry}
\usepackage{color}
\usepackage{array}
\usepackage{booktabs}
\usepackage{amsmath}
\usepackage{amssymb}
\usepackage{graphicx}
\usepackage{setspace}
\usepackage[unicode=true,pdfusetitle,
 bookmarks=true,bookmarksnumbered=false,bookmarksopen=false,
 breaklinks=false,pdfborder={0 0 1},backref=false,colorlinks=false]
 {hyperref}

\makeatletter

\providecommand{\tabularnewline}{\\}

\usepackage{ragged2e}
\usepackage{booktabs}
\usepackage{authblk}
\usepackage{setspace}

\usepackage[parfill]{parskip}
\usepackage{graphicx}
\usepackage{cite}
\usepackage{notoccite}

\title{Nanoscale periodic domain patterns in tetragonal ferroelectrics  a phase-field study}
\author{Ananya Renuka Balakrishna, John E. Huber, Ingo Münch}
\affil{Department of Engineering Science, University of Oxford, Parks Road, Oxford OX1 3PJ, United Kingdom}
\affil{Institute for Structural Analysis, Karlsruhe Institute of Technology, Kaiserstraße 12, 76131 Karlsruhe, Germany}

\date{}                    %% if you don't need date to appear

\makeatother

\begin{document}
\twocolumn[   
\begin{@twocolumnfalse}  
\begin{center}
\textbf{\LARGE{}Nanoscale periodic domain patterns in tetragonal ferroelectrics: }
\par\end{center}{\LARGE \par}

\begin{center}
\textbf{\LARGE{}A phase-field study}
\par\end{center}{\LARGE \par}

\bigskip{}

\begin{center}
Ananya Renuka Balakrishna,$^{a*}$ John E. Huber,$^{a}$ and Ingo
M$\mathrm{\ddot{u}}$nch$^{b}$
\par\end{center}

\bigskip{}

\begin{center}
{\footnotesize{}$^{a}$Department of Engineering Science, University
of Oxford, Parks Road, Oxford OX1 3PJ, England, United Kingdom}
\par\end{center}{\footnotesize \par}

\begin{center}
{\footnotesize{}$^{b}$Institute for Structural Analysis, Karlsruhe
Institute of Technology, Kaiserstraße 12, 76131 Karlsruhe, Germany}
\par\end{center}{\footnotesize \par}

\smallskip{}

\begin{center}
{\scriptsize{}$^{*}$Corresponding author: ananyarb@mit.edu}
\par\end{center}{\scriptsize \par}

\bigskip{}

\bigskip{}

\begin{abstract}
\begin{singlespace}
\textcolor{black}{Ferroelectrics form domain patterns that minimize
their energy subject to imposed boundary conditions. In a linear,
constrained theory, that neglects domain wall energy, periodic domain
patterns in the form of multi-rank laminates can be identified as
minimum-energy states. However, when these laminates (formed in a
macroscopic crystal) comprise domains that are a few nanometers in
size, the domain-wall energy becomes significant, and the behaviour
of laminate patterns at this scale is not known. Here, a phase-field
model, which accounts for gradient energy and strain energy contributions,
is employed to explore the stability and evolution of the nanoscale
multi-rank laminates. The stress, electric field, and domain wall
energies in the laminates are computed. The effect of scaling is also
discussed. In the absence of external loading, stripe domain patterns
are found to be lower energy states than the more complex, multi-rank
laminates, which mostly collapse into simpler patterns. However, complex
laminates can be stabilized by imposing external loads such as electric
field, average strain and polarization. The study provides insight
into the domain patterns that may form on a macroscopic single crystal
but comprising of nanoscale periodic patterns, and on the effect of
external loads on these patterns.}
\end{singlespace}
\end{abstract}
\bigskip{}

\bigskip{}

\end{@twocolumnfalse} ]

\section*{{\small{}Introduction}}

\textcolor{black}{\small{}Domains are regions of a ferroelectric crystal
with uniform polarization. The arrangement, or pattern of these domains
is of great importance in determining the properties of a ferroelectric
at the nanoscale \cite{key-1,key-2,key-3,key-4,key-5,key-6,key-7}.
For example, the effective piezoelectric coefficients depend upon
the fractions of different domain types present \cite{key-6}, and
enhanced actuation can be achieved by using specific domain patterns
\cite{key-7}. Engineered configurations of domains have been used
to make improved actuators, sensors and energy harvesting devices
\cite{key-8}. Domain patterns are particularly important to the working
of nanoscale devices, where relatively few domains are present \cite{key-9,key-10,key-11,key-12}.}{\small \par}

\textcolor{black}{\small{}What determines the pattern of domains?
Where any two domains meet, a domain wall forms, and this has surface
energy proportional to the wall area. But there is also potential
energy, proportional to domain volume, associated with externally
applied stresses and electric fields. Minimization of energy then
produces a competition between the individual energy contributions,
leading to specific patterns. In the absence of external loads, a
single domain state minimizes energy; likewise, for boundary conditions
consistent with a uniform stress or electric field, the single domain
state is an energy minimizer. However, boundary conditions that impose
an average strain or electric displacement can lead to a mixture of
domains. Once a mixture of domains becomes favourable, energy minimization
produces arrangements of domains that relieve internal stresses and
electric fields. Thus \textquotedblleft head-to-tail\textquotedblright{}
polarization arrangements \cite{key-13,key-14,key-15} form across
domain walls, and the spontaneous strains in adjacent domains satisfy
compatibility conditions, well known from the crystallographic theory
of martensite \cite{key-16,key-17}. Where these conditions cannot
be met, the resulting misfit strains cause internal stresses and incompatible
polarization causes electric fields; both produce a contribution to
the net energy. The compatibility rules give a \textquotedblleft constrained
theory\textquotedblright{} that has been widely used to study pattern
formation in ferroelectrics and other materials \cite{key-18,key-19}.
Here, we use this approach to provide a starting point for our simulations. }{\small \par}

\textcolor{black}{\small{}The fact that the domain wall energy scales
with area, while the potential energy scales with volume, introduces
an intrinsic, material specific length scale. Thus energy minimization
can determine both the pattern of domains, and the scale at which
this pattern forms. At a sufficiently coarse scale, the domain wall
energy becomes negligible, and a constrained theory, based on compatibility
alone, is sufficient. However, for the prediction of fine domain patterns,
the contribution of domain wall energy becomes significant. The key
length scale for typical ferroelectrics is of order nanometers, comparable
with the finest spacing of observed domains \cite{key-20,key-21,key-22}.}{\small \par}

\textcolor{black}{\small{}Of course, the materials may not behave
as perfect minimizers of energy. Kinetics has a central role in determining
the ways that domains can evolve. For example, the multiwell nature
of the energy dictates that even if the system is driven towards minimum
energy, it may only reach a local energy minimum. Defects, such as
trapped charge \cite{key-23}, dislocations \cite{key-24}, dopant
atoms \cite{key-25} and so forth, also play a role, affecting both
the energetics and kinetics \cite{key-26}. Nevertheless, substantial
progress in understanding domain patterns has been made by considering
energy minimization in perfect crystals. }{\small \par}

\textcolor{black}{\small{}Domain patterns in single crystal ferroelectrics
commonly take the form of nearly periodic laminations \cite{key-1,key-15,key-27,key-28,key-29,key-30,key-31}.
We can define a rank-1 laminate, as a pattern comprising alternating
layers of two domain types, each of which is a single crystal variant.
Examples include alternating bands of $180^{\circ}$ or $90^{\circ}$
domains. Similarly, simple laminations can themselves be layered together
to produce a higher rank laminate. Based on the linear constrained
theory of laminates, Tsou }\textit{\textcolor{black}{\small{}et al.}}\textcolor{black}{\small{}
\cite{key-18} identified several rank-2 periodic domain patterns
in tetragonal ferroelectrics. Of these, a layered stripe domain pattern
and herringbone domain pattern, are both well-known and commonly observed
\cite{key-13,key-14,key-30,key-32}. Meanwhile, certain other patterns
are rarely seen in experiments \cite{key-33,key-34}, though elements
of these patterns sometimes appear \cite{key-15}. In the present
work, we consider periodic polarization patterns formed in macroscopic
single crystals, but with nanoscale domains. These polarization patterns
are referred to as nanoscale periodic patterns. The theory employed
by Tsou }\textit{\textcolor{black}{\small{}et al.}}\textcolor{black}{\small{}
\cite{key-18} neglected gradient energy due to domain walls and the
elastic/dielectric energy due to disclinations generated at the junctions
of domains. However, these factors could be expected to be of importance,
especially in nanoscale periodic patterns \cite{key-35,key-36}. This
leads us to consider the following questions: Among the periodic laminates
identified by the constrained theory, which are stable when gradient
and elastic energy are considered? Which patterns constitute global
energy minima, and does this depend on the scale? Can imposed states
of average strain or polarization stabilize specific patterns, allowing
the nano-engineering of domain patterns? }{\small \par}

\textcolor{black}{\small{}In the present work, we use a phase-field
approach based on the time-dependent Ginzburg-Landau equations \cite{key-37,key-38,key-39,key-40,key-41}.
This modelling approach is well established for ferroelectrics and
has been successfully applied to study defects \cite{key-42,key-43},
domain structures in thin films \cite{key-44,key-45}, the effect
of flexoelectric coupling on domain patterns \cite{key-46,key-47,key-48,key-49,key-50,key-51,key-52,key-53,key-54},
and microstructural evolution under electro-mechanical loading \cite{key-55}.
The phase-field approach has made a significant contribution in understanding
nanoscale ferroelectrics and reliably accounts for energy contributions
from polarization gradients and misfit strains at domain walls \cite{key-56}.
Several reviews provide a broader appreciation of the capabilities
of the method \cite{key-45,key-57,key-58}. The phase-field simulations
have also been employed to simulate macroscopic ferroelectric properties
by modelling periodic boundary conditions \cite{key-1,key-38,key-59,key-60,key-61,key-62,key-63,key-64}.
For example, systematic studies on the effect of electro-mechanical
boundary conditions on hysteresis, butterfly-loops and polarization
switching have been conducted using periodic conditions \cite{key-38,key-59,key-60,key-61,key-62,key-63}.
While there exists a substantial literature making use of phase field
models to study periodic patterns with stripe-like features in both
2D \cite{key-38,key-59,key-60,key-61,key-62} and 3D \cite{key-63,key-65}
simulations, to date there is relatively little work done to systematically
explore the stability of nanoscale periodic patterns with 2D checkerboard-like
features or with cylindrical domains in 3D that are predicted to form
in tetragonal ferroelectrics \cite{key-27,key-28,key-66}, which is
the goal of the present work. In this paper, the stability of these
complex nanoscale periodic patterns under externally applied strain
and electric field is also explored. }{\small \par}

\textcolor{black}{\small{}The phase-field model \cite{key-42,key-67}
used in the present work enables an initial assessment of the stability
of rank-1 and rank-2 nanoscale periodic domain patterns in tetragonal
ferroelectrics. We then examine the internal fields in the domain
patterns, and consider the effect of scaling on internal energy. Several
of the rank-2 periodic patterns when modelled with nanoscale domains,
collapse into simpler states. However, they can be stabilized by external
loads, and examples of this are given. We also show that, with suitable
loading, simple rank-1 domain patterns can be made to evolve into
more complex, rank-2 patterns, suggesting a mechanism for the formation
of complex domain patterns. Throughout, the phase-field results are
compared with the existing literature on periodic domain patterns. }{\small \par}

\section*{{\small{}Problem description and model}}

\begin{figure}
\begin{centering}
\includegraphics[width=0.9\columnwidth]{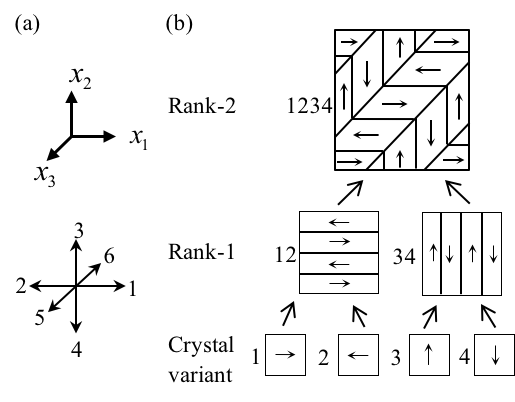}
\par\end{centering}
\caption{\textcolor{black}{\small{}\label{fig:1}(a) Coordinate axes and polarization
directions of the six crystal variants in the tetragonal system. (b)
A rank-2 tree diagram illustrating the domain pattern \textquotedblleft 1234\textquotedblright . }}

\end{figure}

\textcolor{black}{\small{}The objective is to test the stability of
nanoscale periodic patterns that form in a macroscopic single crystal.
To do this, a set of rank-1 and rank-2 periodic laminates identified
by Tsou }\textit{\textcolor{black}{\small{}et al.}}\textcolor{black}{\small{}
\cite{key-18} are modelled using phase-field methods. Following Tsou
}\textit{\textcolor{black}{\small{}et al.}}\textcolor{black}{\small{}
\cite{key-18,key-66} the multi-rank laminates are labelled on the
basis of the crystal variants present in the domain pattern. There
are six crystal variants in the polar tetragonal system, with distinct
spontaneous polarization states as shown in Fig. \ref{fig:1}a. The
structure of a multi-rank laminate consists of layers which are themselves
subdivided into finer layers, the finest laminations comprising material
of a single crystal variant. The arrangement of layers is readily
described using a hierarchical tree structure \cite{key-68}, see
Fig. \ref{fig:1}b. At the lowest level are single crystal variants,
numbered according to their polarization orientation. Ascending the
tree diagram, higher rank laminates are labelled so as to show the
crystal variants present, reading from left to right across the lowest
level of the tree. Thus, in Fig. \ref{fig:1}b, the rank-1 domain
patterns are labelled as \textquotedblleft 12\textquotedblright{}
and \textquotedblleft 34\textquotedblright . These rank-1 domain patterns
can be laminated together to form a rank-2 herringbone domain pattern,
labelled \textquotedblleft 1234\textquotedblright . }{\small \par}

\textcolor{black}{\small{}There are many laminate patterns that can
be formed in this way. However, most are reflections, rotations or
inversions of other patterns. Taking out these symmetrical copies,
Tsou }\textit{\textcolor{black}{\small{}et al.}}\textcolor{black}{\small{}
\cite{key-18} found just two distinct laminates of rank-1 and eight
of rank-2. We model nanoscale periodic cells of these patterns and
seek equilibrium states. }{\small \par}

\textcolor{black}{\small{}A phase-field model developed by Landis
and co-workers \cite{key-42,key-67} is used. This model has been
calibrated using the material properties of barium titanate (BaTiO$_{3}$)
in its tetragonal phase \cite{key-67}. The order parameter is the
local polarization $P_{i}$, and the polarization field is allowed
to approach equilibrium by relaxation using the Ginzburg-Landau equation
\cite{key-67}:}{\small \par}

\textcolor{black}{\small{}
\begin{equation}
\left(\frac{\partial\psi}{\partial P_{i,j}}\right)_{,j}-\frac{\partial\psi}{\partial P_{i}}=\beta\dot{P_{i}},\label{eq:1}
\end{equation}
}{\small \par}

\textcolor{black}{\small{}Here $\psi$ is the Helmholtz free energy
per unit volume. An arbitrary polarization viscosity, $\beta$, is
introduced for numerical purposes, and is controlled as a relaxation
parameter. Equilibrium states satisfy Eq. \ref{eq:1} with $\beta=0$.
The Helmholtz free energy is described by: }{\small \par}

\textcolor{black}{\footnotesize{}
\begin{align}
\psi & =\frac{1}{2}a_{ijkl}P_{i,j}P_{k,l}+\frac{1}{2}\overline{a}_{ij}P_{i}P_{j}+\frac{1}{4}\overline{\overline{a}}_{ijkl}P_{i}P_{j}P_{k}P_{l}\nonumber \\
 & ~+\frac{1}{6}\overline{\overline{\overline{a}}}_{ijklmn}P_{i}P_{j}P_{k}P_{l}P_{m}P_{n}+\frac{1}{8}\overline{\overline{\overline{\overline{a}}}}_{ijklmnrs}P_{i}P_{j}P_{k}P_{l}P_{m}P_{n}P_{r}P_{s}\nonumber \\
 & ~+b_{ijkl}\epsilon_{ij}P_{k}P_{l}+\frac{1}{2}c_{ijkl}\epsilon_{ij}\epsilon_{kl}+\frac{1}{2}f_{ijklmn}\epsilon_{ij}\epsilon_{kl}P_{m}P_{n}\nonumber \\
 & \ +\frac{1}{2}g_{ijklmn}\epsilon_{ij}P_{k}P_{l}P_{m}P_{n}+\frac{1}{2\kappa_{0}}(D_{i}-P_{i})(D_{i}-P_{i})\label{eq:2}
\end{align}
}{\footnotesize \par}

\textcolor{black}{\small{}where $D_{i}$ is the electric displacement,
$\epsilon_{ij}$ is the strain tensor and $\kappa_{0}$ is the permittivity
of free space. The tensorial coefficients $\mathbf{a},\mathbf{b,}\mathbf{c,f}$
and $\mathbf{g}$ are given, along with further details of the model
and the material properties of BaTiO$_{3}$ at room temperature in
the work by Landis and co-workers \cite{key-42,key-67}. These parameters
were chosen to represent the multiwell free energy and provide the
correct symmetries to model the material properties in the tetragonal
phase. In Eq. \ref{eq:2}, the energy contribution from surface energy
is assumed to be negligible since the work focusses on macroscopic
crystals. The phase-field problem is solved using finite element methods,
with 8-noded quadratic elements in two dimensional (2D) simulations
or brick elements in three dimensional (3D) simulations. An element
size of 1nm was chosen, such that domain walls typically span over
about three elements. }{\small \par}

\begin{figure}
\begin{centering}
\includegraphics[width=0.9\columnwidth]{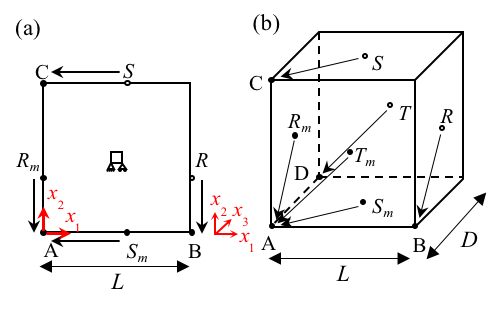}
\par\end{centering}
\caption{\textcolor{black}{\small{}\label{fig:2}Schematic representation of
the periodic cell and boundary conditions on (a) a square of side
$L$ (b) a cuboid of depth $D$. Arrows indicate the nodal relations
in periodic boundary conditions. }}
\end{figure}

\textcolor{black}{\small{}The material model is fully 3D, but several
of the domain patterns considered have polarization in a single plane
with all domain walls perpendicular to that plane. In these cases
the prismatic nature of the patterns is exploited allowing them to
be modelled in two dimensions, with plane strain and electric field
conditions. This is practically expedient because the simulations
are computationally intensive. However, it does limit the simulation
by preventing out-of-plane polarization. Our experience with a wider
range of simulations indicates that this restriction does not greatly
affect the model outcomes. Domain patterns with out-of-plane polarizations
are modelled in 3D. To model a regular repeating laminate structure,
periodic boundary conditions are imposed, see Fig. \ref{fig:2}. A
periodic square of side$L=40$nm, Fig. \ref{fig:2}a, and a periodic
cuboid with depth $D~<~L$, Fig. \ref{fig:2}b, are modelled in the
2D and 3D simulations respectively. These periodic cells when repeated
infinitely form a macroscopic object like a thin film or a single
crystal. }{\small \par}

\textcolor{black}{\small{}The mid-elements of the periodic cell shown
in Fig. \ref{fig:2}(a-b) have simple supports, while periodic conditions
are enforced on the boundary nodes, controlling the displacement  $u_{i}$,
electric potential $\phi$, and polarization $P_{i}$. For a typical
node $R$, these conditions are: }{\small \par}

\textcolor{black}{\footnotesize{}
\begin{align}
u_{i}(L,x_{2},x_{3})\mid_{R}-u_{i}(L,0,0)\mid_{\mathrm{B}} & =u_{i}(0,x_{2},x_{3})\mid_{R_{m}}-u_{i}(0,0,0)\mid_{\mathrm{A}},\nonumber \\
\phi(L,x_{2},x_{3})\mid_{R} & =\phi(0,x_{2},x_{3})\mid_{R_{m}},\nonumber \\
P_{i}(L,x_{2},x_{3})\mid_{R} & =P_{i}(0,x_{2},x_{3})\mid_{R_{m}}\qquad i=1,n.\label{eq:3}
\end{align}
}{\footnotesize \par}

\textcolor{black}{\small{}Here $n$ is the number of dimensions, and
for 2D simulations $x_{3}=0$. The notation \textquoteleft $\mid_{X}$\textquoteright{}
is used to indicate a value at a node \textquoteleft $X$\textquoteright{}
as labelled in Fig. \ref{fig:2}. Similar conditions apply to typical
node $S$, and (in the 3D case) $T$. Care is needed at corners in
2D and at edges in 3D to avoid duplication of boundary conditions.
These boundary conditions allow the domain pattern to adopt periodic
strain fields $\epsilon_{ij}(x_{1},x_{2},x_{3})$ with zero average
stress. That is, the strain fields $\epsilon_{ij}$ are dependent
on the relative displacement between the corner nodes, for example
node \textquoteleft A\textquoteright{} and node \textquoteleft B\textquoteright ,
which are not bound. These corner nodes are free to adopt any displacement
values $u_{i}$ such that the periodic cell possesses zero average
stress. Meanwhile, periodic values of electric potential $\phi$ with
zero average electric field, and polarization $P_{i}$ are imposed
on the boundary nodes. We shall later refer to the boundary conditions
represented by Eq. \ref{eq:3} as \textquotedblleft zero external
load\textquotedblright{} conditions. Eq. \ref{eq:3} imposes the boundary
conditions of a bulk material and previous literature \cite{key-3,key-14,key-15,key-30,key-32,key-69,key-70,key-71}
indicates that periodic domain patterns have been experimentally observed
in bulk samples. }{\small \par}

\textcolor{black}{\small{}The phase-field simulation is initialized
by imposing a polarization field consistent with the repeating unit
of a laminate pattern on the periodic cell. The nodal polarization
values are set equal to the spontaneous polarization of the domain,
producing sharp discontinuities at domain walls in the initial state.
The displacement and electric potential values are initialized at
zero. Note that this initial state may not satisfy electro-mechanical
compatibility conditions, however it provides a state of the order
parameter $P_{i}$ that approximates an equilibrium pattern. }{\small \par}

\textcolor{black}{\small{}During the early stages of each simulation,
high values of polarization viscosity, $\beta$, are used to facilitate
the resolution of domain walls into a continuous polarization field
and displacement field. The sharp discontinuity at the domain walls
rapidly relaxes, but the overall pattern of polarization is otherwise
almost unchanged at this stage. Once this initial settling has occurred
\textendash{} a condition that will be referred to as the \textquoteleft settled
state\textquoteright{} \textendash{} the free energy of the pattern,
though not necessarily yet in equilibrium, is calculated. Subsequently,
larger relaxation steps are used, allowing the simulation to evolve
towards equilibrium, with the possibility of changing the polarization
pattern in the process. }{\small \par}

\section*{{\small{}Results and Discussion}}

\textcolor{black}{\small{}Before presenting the results of the simulations,
several normalizations are introduced. The Helmholtz free energy per
unit volume is normalized as, $\tilde{\psi}=(\psi_{0}-\psi)/\psi_{0}$,
where $\psi_{0}$ corresponds to the energy per unit volume of a spontaneously
polarized monodomain. Here $\psi_{0}<0$ , while $\psi=0$ corresponds
to the cubic state with  $P_{i}=0$. The polarization and strains
are normalized by $P_{0}=0.26\mathrm{C/m^{2}}$ and $\epsilon_{0}=0.0082$
respectively, which arises from the spontaneous state of barium titanate.
Note that BaTiO$_{3}$ has tetragonal symmetry, and possesses a transverse
spontaneous strain $\epsilon_{0}^{t}=-0.0027$. $E_{0}=21.82\mathrm{MV/m}$
is the normalization constant for electric field, and corresponds
to the critical field required to cause homogeneous $180^{\circ}$
switching of a spontaneously polarized monodomain. Equating mechanical
and electrical energies, the normalization for stress is derived as
$\sigma_{0}=E_{0}P_{0}/\epsilon_{0}=692\mathrm{MPa}$. The characteristic
length scale in this model is $l_{0}=\sqrt{a_{0}P_{0}/E_{0}}=1\mathrm{nm}$,
where $a_{0}=1\times10^{-10}\mathrm{Vm^{3}/C}$ is a coefficient used
in specifying the gradient energy term in Eq. \ref{eq:2} \cite{key-67}.
The electric potential normalization is derived from the electric
field and length scale normalizations as $\phi_{0}=E_{0}l_{0}=0.022\mathrm{V}$.}{\small \par}

\subsection*{{\footnotesize{}Rank-1 domain patterns}}

\begin{figure}
\begin{centering}
\includegraphics[width=1\columnwidth]{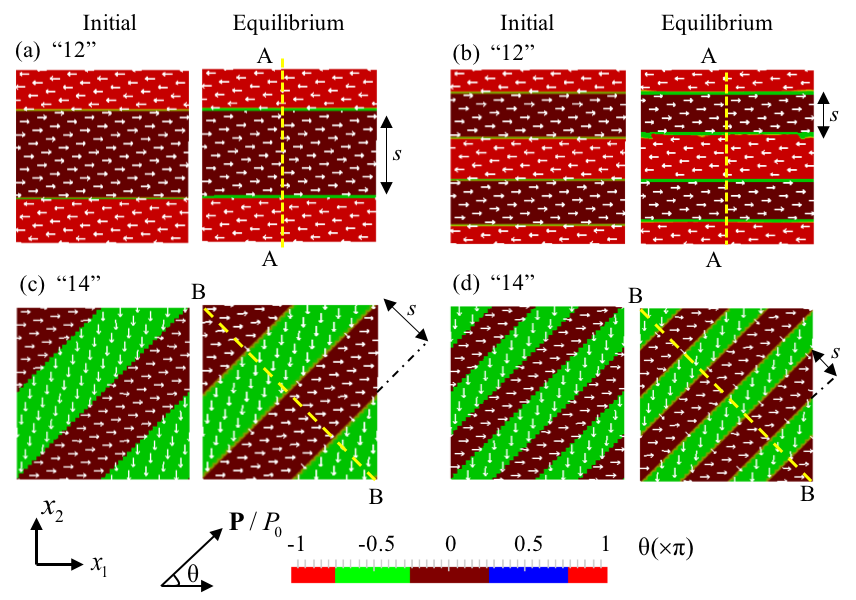}
\par\end{centering}
\caption{\textcolor{black}{\small{}\label{fig:3}Testing the stability of rank-1
domain patterns under zero external load conditions. $\mathbf{P}/P_{0}$
values in (a) laminate \textquotedblleft 12\textquotedblright{} with
$s=20\mathrm{nm}$, (b) laminate \textquotedblleft 12\textquotedblright{}
with $s=10\mathrm{nm}$, (c) laminate \textquotedblleft 14\textquotedblright{}
with $s=14.1\mathrm{nm}$, and (d) laminate \textquotedblleft 14\textquotedblright{}
with $s=7.1\mathrm{nm}$. AA and BB indicate cross-sections perpendicular
to the domain walls. }}
\end{figure}

\textcolor{black}{\small{}At first, we test the stability of the rank-1
domain patterns under zero external load conditions as given by Eq.
\ref{eq:3}. The two rank-1 domain patterns identified in tetragonal
ferroelectrics are the \textquotedblleft 12\textquotedblright{} laminate
pattern with alternating $180^{\circ}$ domain bands and the \textquotedblleft 14\textquotedblright{}
laminate pattern with alternating $90^{\circ}$ domain stripes, see
Fig. \ref{fig:3}. The \textquotedblleft 12\textquotedblright{} domain
pattern is modelled with two different domain spacings $s=20\mathrm{nm}$
and $10\mathrm{nm}$ defined along cross-section AA, see Fig. \ref{fig:3}(a-b).
Similarly the \textquotedblleft 14\textquotedblright{} domain pattern
is modelled at two spacings, with $s=14.1\mathrm{nm}$ and $7.1\mathrm{nm}$
defined along cross-section BB, as shown in Fig. \ref{fig:3}(c-d). }{\small \par}

\textcolor{black}{\small{}At equilibrium, the sharp domain walls in
the initial states become smooth, but the patterns are otherwise unchanged,
see Fig. \ref{fig:3}(a-d). Nanoscale periodic patterns with bundles
of stripe-like or band-like features have been observed to be stable
in experiments on BaTiO$_{3}$ \cite{key-27} for example by McGilly
}\textit{\textcolor{black}{\small{}et al.}}\textcolor{black}{\small{}
\cite{key-15} and Schilling }\textit{\textcolor{black}{\small{}et
al.}}\textcolor{black}{\small{} \cite{key-29,key-70}. This stability
of nanoscale patterns with $180^{\circ}$ or $90^{\circ}$ domains
is in agreement with the phase-field simulation results, see Fig.
\ref{fig:3}(a-d). }{\small \par}

\begin{figure}
\begin{centering}
\includegraphics[width=1\columnwidth]{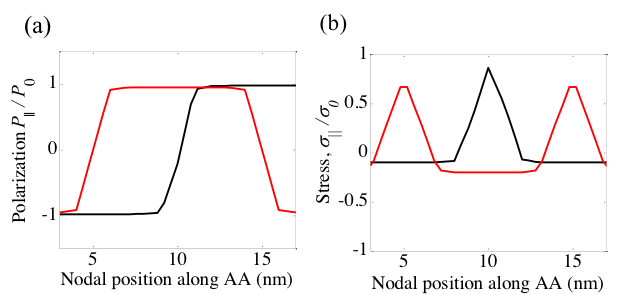}
\par\end{centering}
\textcolor{black}{\small{}\caption{\textcolor{black}{\small{}\label{fig:4}Nodal values of (a) polarization
$P_{||}$, parallel to the domain wall and (b) stress component $\sigma_{||}$,
parallel to the domain wall on cross-section AA in the $180^{\circ}$
band domain pattern. }}
}{\small \par}

\end{figure}

\textcolor{black}{\small{}The variations in polarization and stress
developed on section AA in the $180^{\circ}$ domain pattern \textquotedblleft 12\textquotedblright{}
are shown in Fig. \ref{fig:4}(a-b). From Fig. \ref{fig:4}a, the
domain wall width is close to 2nm, consistent with the value for an
isolated domain wall given by Su and Landis \cite{key-67}, and is
unaffected by changing the domain spacing from 10nm to 20nm. These
values are also consistent with the results from the quasi-one-dimensional
analysis of tetragonal twins by Cao and Cross \cite{key-72} and match
separate calculations by Hlinka and Márton \cite{key-73}. Away from
the domain walls, the polarization magnitude is close to the spontaneous
value (within 1\%) and there is no polarization perpendicular to the
walls. The electric field is everywhere negligible. The domain wall
width in ferroelectrics can also be influenced by surface effects
and elastic interaction between nonferroelastic domain walls, which
are discussed in previous investigations \cite{key-74,key-75}.}{\small \par}

\textcolor{black}{\small{}}
{\small \par}

\textcolor{black}{\small{}At equilibrium, stresses develop parallel
to the $180^{\circ}$ walls as shown in Fig. \ref{fig:4}b, while
stresses perpendicular to the walls are zero. Here, these stresses
correspond to the values of $\sigma_{11}$ and $\sigma_{22}$ respectively.
With spacing $s=20\mathrm{nm}$ the tensile stress peaks at about
$0.85\sigma_{0}$ in the domain walls, balanced by compressive stresses
of about $-0.1\sigma_{0}$ in the domains. On decreasing the spacing
to $s=10\mathrm{nm}$, the peak stress reduces to about $0.65\sigma_{0}$,
while the compressive stresses in the domains double in magnitude,
becoming $-0.2\sigma_{0}$. The energy per unit area associated with
the presence of domain walls was calculated using }{\small \par}

\textcolor{black}{\small{}
\begin{equation}
\gamma_{\mathrm{w}}=\frac{\int_{V}(\psi-\psi_{0})\mathrm{d}V}{A_{\mathrm{w}}}\label{eq:4}
\end{equation}
}{\small \par}

\textcolor{black}{\small{}where $A_{\mathrm{w}}$ is the wall area
within the periodic cell of volume $V$. We found $\gamma_{\mathrm{w}}\sim13.1\mathrm{mJ/m^{2}}$
when $s=20\mathrm{nm}$ and $\gamma_{\mathrm{w}}\sim12.6\mathrm{mJ/m^{2}}$
for $s=10\mathrm{nm}$. This compares with the value of $14.8\mathrm{mJ/m^{2}}$
for the case of an isolated domain wall $s\rightarrow\infty$ \cite{key-67}.
It is interesting to note that the domain wall energy in the periodic
laminate is less than the energy of an isolated domain wall, and dependent
upon the domain spacing. The decrease in stresses at the domain walls
accounts for the difference. Similar magnitudes of domain wall energy
have been found in previous investigations \cite{key-76}. }{\small \par}

\begin{figure}
\begin{centering}
\includegraphics[width=1\columnwidth]{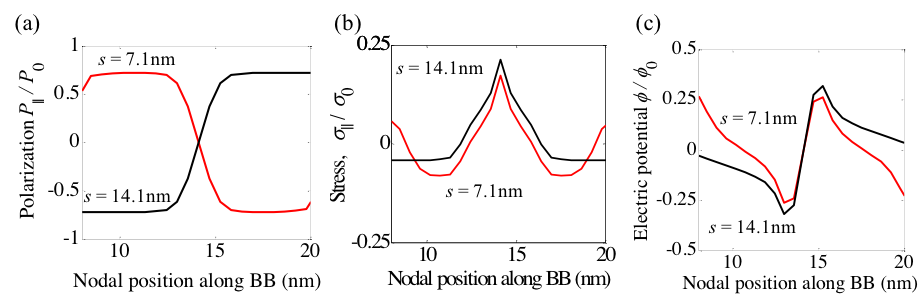}
\par\end{centering}
\caption{\textcolor{black}{\small{}\label{fig:5}Nodal values of (a) polarization
, parallel to the domain wall and (b) stress component and (c) electric
potential on cross-section BB in the stripe domain pattern. }}
\end{figure}

\textcolor{black}{\small{}The polarization variation on section BB
in the $90^{\circ}$ stripe domain pattern \textquotedblleft 14\textquotedblright ,
with spacing of 14.1nm and 7.1nm, is shown in Fig. \ref{fig:5}a.
Away from the domain walls, the polarization magnitude approaches
$P_{0}$ with components of magnitude $P_{0}/\sqrt{2}$ parallel and
perpendicular to the domain walls. From Fig. \ref{fig:5}a the $90^{\circ}$
domain wall width is \textasciitilde{}3nm, in agreement with Su and
Landis \cite{key-67}.}{\small \par}

\textcolor{black}{\small{}At equilibrium, tensile stresses parallel
to the $90^{\circ}$ domain walls develop, balanced by compressive
stresses in the domains, see Fig. \ref{fig:5}b. The stresses at the
domain walls peak at about $0.2\sigma_{0}$ for $s=14.1\mathrm{nm}$
and reduce to $0.17\sigma_{0}$ when $s=7.1\mathrm{nm}$. This differs
from the value of $\sim0.33\sigma_{0}$ calculated for the case $s\rightarrow\infty$
\cite{key-67}; the difference can be attributed to the fine spacing
of the domains, which places the domains into significant compression
and hence affects the stress distribution in the domain walls. The
domain wall energies are calculated to be $\sim6.0\mathrm{mJ/m^{2}}$
for $s=14.1\mathrm{nm}$ and $\sim5.8\mathrm{mJ/m^{2}}$ for $s=7.1\mathrm{nm}$.
These values are consistent with the energy calculations by Su and
Landis \cite{key-67}. The stresses developed in the $90^{\circ}$
stripe domain pattern are rather low in comparison to the $180^{\circ}$
band domain pattern, and this accounts for the difference in the $90^{\circ}$
and $180^{\circ}$ domain wall energies. }{\small \par}

\textcolor{black}{\small{}Non-zero local electric fields are developed
at the $90^{\circ}$ domain walls, as shown by the change in electric
potential values in Fig. \ref{fig:5}c. Note that the pattern has
a non-zero average polarization. The electric fields are in the direction
opposite to this net polarization, and cause deviation from the spontaneously
polarized state at the walls. A consequence is that there are electric
fields in the domains that are aligned with the average polarization.
With $s=14.1\mathrm{nm}$ the change in electric potential across
a domain wall is about $\Delta\phi\sim0.6\phi_{0}$, corresponding
to an electric field of magnitude $\sim0.2E_{0}$. On decreasing the
spacing to $s=7.1\mathrm{nm}$, $\Delta\phi$ drops to $\sim0.5\phi_{0}$
with the corresponding electric field magnitude reduced to $\sim0.17E_{0}$.
However, the electric field in the polarized domains increases with
decrease in spacing. }{\small \par}

\textcolor{black}{\small{}We also tested whether keeping the cell
size the same but changing the relative position of the domain walls
affects the overall energy. If moving one domain wall within the cell
affects the energy then there could be a minimum energy arrangement
\textendash{} effectively a \textquotedblleft preferred\textquotedblright{}
separation distance between domain walls. We found that this is not
the case: the domain walls in the cell were in neutral equilibrium
over a wide range of separation distances. Details of these simulations
are omitted for brevity. The finding is consistent with observations
that commonly show an approximately periodic pattern, but with considerable
variation in distances between domain walls \cite{key-14,key-34}.
This is significant because it indicates that perfectly engineered
domain configurations with regular spacing are unlikely to form naturally
and will require special conditions to enforce regularity. An example
of such conditions is given later in section III-C. }{\small \par}

\subsection*{{\footnotesize{}Rank-2 domain patterns}}

\begin{figure}
\begin{centering}
\includegraphics[width=1\columnwidth]{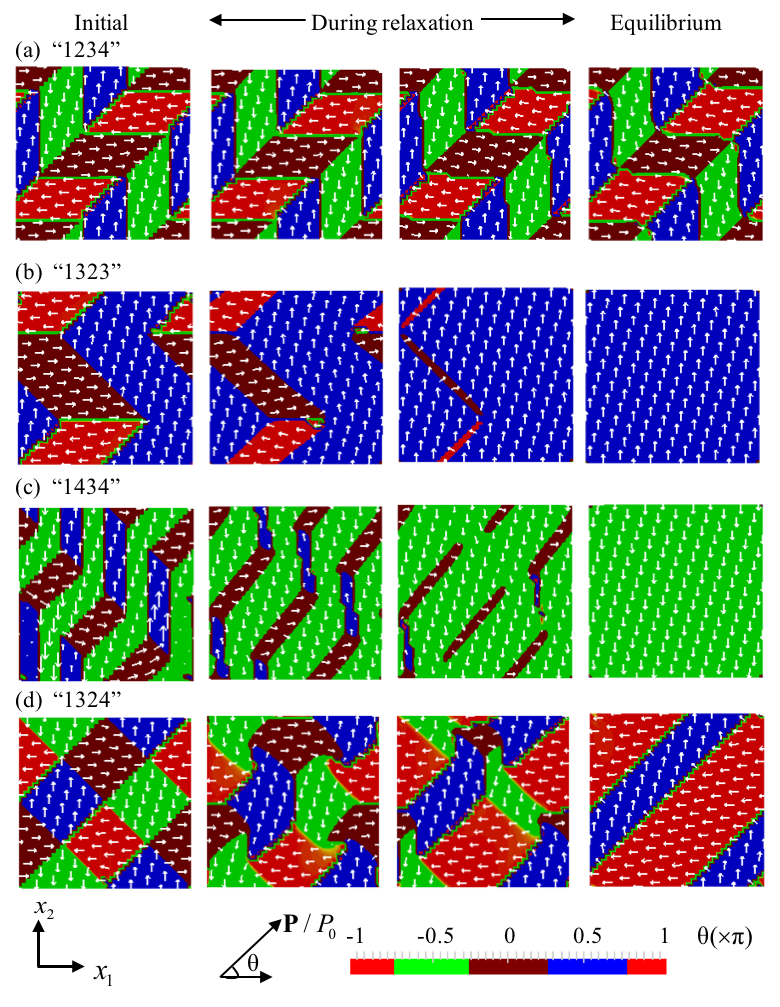}
\par\end{centering}
\caption{\textcolor{black}{\small{}\label{fig:6}Testing the stability of rank-2
domain patterns with in-plane polarizations under zero external load
conditions, laminates: (a) \textquotedblleft 1234\textquotedblright{}
(b) \textquotedblleft 1323\textquotedblright{} (c) \textquotedblleft 1434\textquotedblright{}
and (d) \textquotedblleft 1324\textquotedblright . Sub-figures map
domain rearrangements during relaxation from the initial state (far
left) to the final equilibrium state (far right). }}
\end{figure}

\textcolor{black}{\small{}Next we test the stability of rank-2 periodic
domain patterns under the zero external load conditions. Among the
rank-2 laminates identified by Tsou }\textit{\textcolor{black}{\small{}et
al.}}\textcolor{black}{\small{} \cite{key-18}, four domain patterns
which contain polarization only in a single plane with domain walls
perpendicular to that plane are first tested, see Fig. \ref{fig:6}(a-d).
All of the rank-2 domain patterns shown in Fig. \ref{fig:6}(a-d)
are modelled with zero average stress and electric field conditions
as described by Eq. \ref{eq:3}. Here, note that the strain fields
$\epsilon_{ij}$ adopted by the periodic cell vary based on the type
of domain pattern imposed in the initial state. The volume average
free energy $\tilde{\psi}$, average axial components of strain, $\tilde{\epsilon}_{ij}$,
and polarization, $\tilde{P}_{i}$, adopted by these domain patterns
in their settled states are given in Table I. Each pattern was simulated
in a square region of side $L=40\mathrm{nm}$ using plane strain and
electric field conditions. }{\small \par}

\textcolor{black}{\small{}The herringbone domain pattern \textquotedblleft 1234\textquotedblright{}
shown in Fig. \ref{fig:6}a, is found to be stable at this scale.
This domain pattern contains several domain walls in the periodic
cell, which accounts for its relatively high free energy in comparison
to the rank-1 domain patterns. }{\small \par}

\textcolor{black}{\small{}Stresses in the herringbone domain pattern
are developed at the $180^{\circ}$ domain walls, while non-zero local
electric fields are observed at the $90^{\circ}$ domain walls. During
relaxation the domain walls drifted slightly, resulting in a small
magnitude of net polarization $(\sim0.05P_{0})$ due to the slightly
unequal sizes of the $\pm P_{2}$ domains at equilibrium. Herringbone
patterns are observed to be stable in ferroelectrics \cite{key-14,key-27,key-30,key-77},
thereby confirming the phase-field results in Fig. \ref{fig:6}a. }{\small \par}

\begin{figure}
\begin{centering}
\includegraphics[width=1\columnwidth]{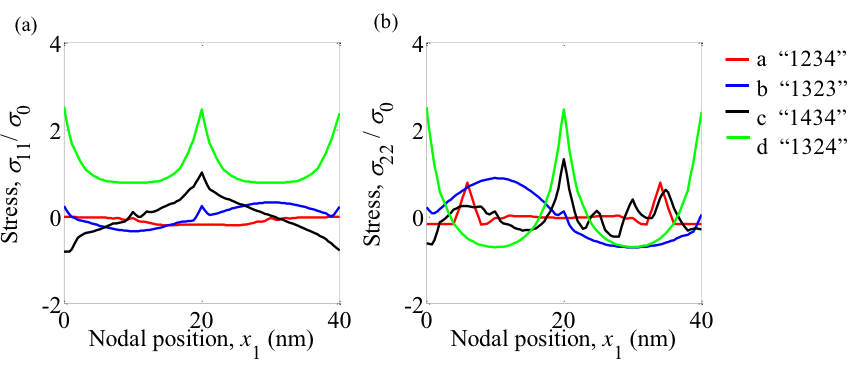}
\par\end{centering}
\caption{\textcolor{black}{\small{}\label{fig:7}Stress distributions in the
rank-2 periodic domain patterns with in-plane polarizations: (a) $\sigma_{11}$
(b) $\sigma_{22}$ at a cross-section $x_{2}=L/2$, at varying $x_{1}$
positions. }}

\end{figure}

\textcolor{black}{\small{}The periodic domain pattern in Fig. \ref{fig:6}b
shows laminate \textquotedblleft 1323\textquotedblright{} which contains
alternating bands of $\pm P_{1}$ $180^{\circ}$ domains interspersed
with a zig-zag stripe which is a single $+P_{2}$ domain. This domain
pattern was identified as a minimum energy state in the linear constrained
theory\cite{key-18}; that is, all domain walls satisfy compatibility
conditions. However, when \textquotedblleft 1323\textquotedblright{}
is modelled with nanoscale domains, the pattern is found to be unstable:
during relaxation it dissolves to form a uniformly polarized $+P_{2}$
domain. At the settled state i.e., where domain walls are resolved
but the polarization pattern is unchanged, the $\pm P_{1}$ domains
experience a tensile stress $\sigma_{22}$, while the $P_{2}$ domains
experience compressive stress $\sigma_{22}$. This is to be expected
because the spontaneous strains component $\epsilon_{22}$ differs
between the $\pm P_{1}$ domains (where it is $\epsilon_{0}^{t}$)
and the domains (where it is $\epsilon_{0}$). Since these vertical
zig-zag stripes are assumed to be perfectly adhered to each other,
misfit stress arises from the difference in spontaneous strain. Fig. \ref{fig:7}
shows the direct stresses $\sigma_{11},\sigma_{22}$ evaluated at
a cross-section of the model where $x_{2}=L/2$. It is clear that
the herringbone pattern (label \textquoteleft a\textquoteright ) has
relatively low values of stress throughout, while the tensile and
compressive stresses in the alternating bands of the \textquotedblleft 1323\textquotedblright{}
domain pattern (label \textquoteleft b\textquoteright ) are of order
$\sigma_{0}$. This contribution of elastic energy due to misfit stress
explains the lack of stability and accounts for the net energy of
the domain pattern, see Table I. }{\small \par}

\textcolor{black}{\small{}An additional factor contributing to the
high energy of the \textquotedblleft 1323\textquotedblright{} domain
pattern is the presence of disclinations at the junctions of domains.
These arise because the unstressed condition of a $90^{\circ}$ domain
wall produces a slight rotation of the crystal lattice, dependent
on the tetragonal $c:a$ ratio. In barium titanate this rotation is
about $0.63^{\circ}$. Upon circulating a continuous junction of domains,
no net rotation of the crystal lattice is permissible. Consequently
the domains are stressed if the set of domain walls at a junction
would otherwise produce a net rotation. For example, the domains in
the \textquotedblleft 1234\textquotedblright{} herringbone pattern
are so arranged that there is no net lattice rotation upon circulating
any junction: opposite rotations on passing $90^{\circ}$ domain walls
cancel. However, in the \textquotedblleft 1323\textquotedblright{}
pattern, circulation of a domain junction implies two consecutive
lattice rotations of $0.63^{\circ}$, resulting in a disclination. }{\small \par}

\begin{table}
\noindent %
\begin{tabular*}{1\columnwidth}{@{\extracolsep{\fill}}>{\raggedright}p{0.15\columnwidth}>{\centering}p{0.2\columnwidth}cccc}
\toprule 
{\scriptsize{}Rank \textendash{} 2 periodic laminates} &
{\scriptsize{}Average free energy } &
\multicolumn{2}{>{\centering}p{0.2\columnwidth}}{{\scriptsize{}Average strain components $\tilde{\epsilon}_{ij}=\left\langle \epsilon_{ij}\right\rangle /\epsilon_{0}$}} &
\multicolumn{2}{>{\centering}p{0.28\columnwidth}}{{\scriptsize{}Average polarization components $\tilde{P}_{i}=\left\langle P_{i}\right\rangle /P_{0}$}}\tabularnewline
 &
{\scriptsize{}$\tilde{\psi}$} &
{\scriptsize{}$\tilde{\epsilon}_{11}$} &
{\scriptsize{}$\tilde{\epsilon}_{22}$} &
{\scriptsize{}$\tilde{P}_{1}$} &
{\scriptsize{}$\tilde{P}_{2}$}\tabularnewline
\midrule
{\scriptsize{}(a) ``1234''} &
{\scriptsize{}0.49} &
{\scriptsize{}0.29} &
{\scriptsize{}0.29} &
{\scriptsize{}$-$0.05} &
{\scriptsize{}0.05}\tabularnewline
{\scriptsize{}(b) ``1323''} &
{\scriptsize{}0.48} &
{\scriptsize{}0.32} &
{\scriptsize{}0.29} &
{\scriptsize{}0.47} &
{\scriptsize{}0.00}\tabularnewline
{\scriptsize{}(c) ``1434''} &
{\scriptsize{}0.77} &
{\scriptsize{}0.06} &
{\scriptsize{}0.54} &
{\scriptsize{}0.24} &
{\scriptsize{}$-$0.24}\tabularnewline
{\scriptsize{}(d) ``1324''} &
{\scriptsize{}0.75} &
{\scriptsize{}0.28} &
{\scriptsize{}0.28} &
{\scriptsize{}0.00} &
{\scriptsize{}0.00}\tabularnewline
\bottomrule
\end{tabular*}

\caption{\textcolor{black}{\small{}\label{tab:1}The normalized volume average
free energy $\tilde{\psi}$, average axial strains $\tilde{\epsilon}_{ij}$,
and polarization $\tilde{P}_{i}$ adopted by rank-2 periodic domain
patterns with in-plane polarization in their settled state. }}
\end{table}

\textcolor{black}{\small{}The domain rearrangement process during
relaxation in Fig. \ref{fig:6}b, indicates the growth of domains
with a larger initial volume fraction, i.e., $P_{2}$ domains, in
comparison to the $\pm P_{1}$ domains which reduce in size. The relaxation
process eventually eliminates the misfit stresses of the \textquotedblleft 1323\textquotedblright{}
domain pattern, resulting in a uniformly polarized, stress-free domain
at equilibrium. }{\small \par}

\textcolor{black}{\small{}The domain pattern \textquotedblleft 1434\textquotedblright{}
shown in Fig. \ref{fig:6}c contains stripe-like and herringbone-like
domains. This domain pattern is unstable at the scale simulated ($L=40\mathrm{nm}$),
dissolving into a monodomain. During relaxation, the $-P_{2}$ domains,
which have the largest volume fraction, grow relative to neighbouring
domains. Referring to Fig. \ref{fig:7}, the \textquotedblleft 1434\textquotedblright{}
pattern has stressed domains which contribute to an increased free
energy. }{\small \par}

\textcolor{black}{\small{}The checkerboard domain pattern \textquotedblleft 1324\textquotedblright{}
contains repeating groups of $90^{\circ}$ domains forming closed
polarization loops, or vortices, see Fig. \ref{fig:6}d. This pattern
too is unstable: it dissolves into a rank-1 stripe domain pattern
at equilibrium. As noted by Tsou }\textit{\textcolor{black}{\small{}et
al.}}\textcolor{black}{\small{} \cite{key-18} the checkerboard pattern
contains the strongest disclinations among the four in-plane rank-2
domain patterns of Fig. \ref{fig:6}(a-d), leading to stresses of
order $2\sigma_{0}$, and hence high energy. This can clearly be observed
in the stress distributions of Fig. \ref{fig:7}. During relaxation,
the domain junctions dissolve by formation of $180^{\circ}$ domain
walls. Symmetry breaking occurs and the final state is a rank-1 domain
pattern with non-zero average polarization at equilibrium. }{\small \par}

\textcolor{black}{\small{}Although the herringbone pattern \textquotedblleft 1234\textquotedblright{}
is well known both at the nano- and microscales, the other patterns
shown in Fig. \ref{fig:6}(b-d) have relatively rarely been observed
and reported \cite{key-27,key-18}. A recent work by Tang }\textit{\textcolor{black}{\small{}et
al.}}\textcolor{black}{\small{} \cite{key-28} reports a periodic
array of flux-closure domains in PbTiO$_{3}$. However, this polarization
pattern differs from the checkerboard pattern in Fig. \ref{fig:6}d,
in that the observed pattern possesses a $180^{\circ}$ domain wall
at the vortex core. This domain pattern observed by Tang }\textit{\textcolor{black}{\small{}et
al.}}\textcolor{black}{\small{} is referred in section III-C of this
paper, where the phase-field simulations indicate its stability in
the presence of external loads. }{\small \par}

\textcolor{black}{\small{}So far, the experimental observations on
nanoscale periodic polarization patterns available in literature are
in agreement with our phase-field simulations. However, it is of interest
to consider how scaling of the periodic cell size affects the energy
density of these patterns. Crudely, we can think of the free energy
$\int\psi\mathrm{d}V$ as including a contribution due to gradient
energy $\int\psi_{\mathrm{w}}\mathrm{d}V$ and an elastic/dielectric
energy part $\int\psi_{e}\mathrm{d}V$. The gradient energy is mainly
due to domain walls and scales with domain wall area. Meanwhile the
contributions to the elastic/dielectric energy arise from misfit stress
and disclinations, with disclinations being the main factor. For the
$i$th disclination in a given pattern, the energy contribution scales
with the square of its disclination angle, $\theta_{i}$, the elastic
modulus, $c\sim67\mathrm{GPa}$ \cite{key-78}, and the volume $V_{i}$
of material in proximity to the disclination (taken here to be the
set of material points closest to the $i$th disclination). Note the
laminates in Fig. \ref{fig:6}(b-d) contain disclination dipoles.
For example, consider laminate \textquotedblleft 1324\textquotedblright{}
in Fig. \ref{fig:6}d \textendash{} in this pattern, alternating positive
and negative disclinations are observed upon circulating domain junctions.
The long-range elastic field induced by these disclination dipoles
cancel each other and are neglected in the present approximation.
A rough estimate of the free energy can then be calculated using, }{\small \par}

\textcolor{black}{\small{}
\begin{equation}
\int\psi(\mathbf{x})\mathrm{d}V\sim\int_{A_{90}}\gamma_{90}\mathrm{d}A_{90}+\int_{A_{180}}\gamma_{180}\mathrm{d}A_{180}+\underset{i}{\sum}c\theta_{i}^{2}V_{i}.\label{eq:5}
\end{equation}
}{\small \par}

\begin{figure}
\begin{centering}
\includegraphics[width=0.6\columnwidth]{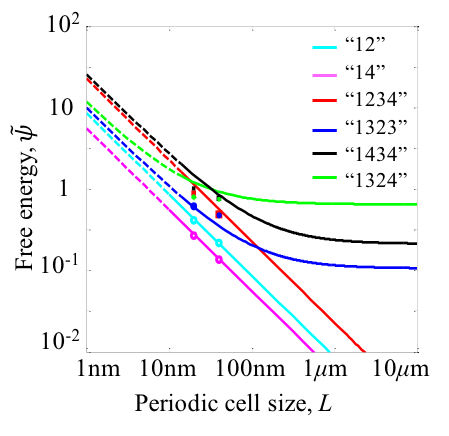}
\par\end{centering}
\caption{\textcolor{black}{\small{}\label{fig:8}Estimates of the variation
of free energy for rank-1 and rank-2 domain patterns as a function
of periodic cell size, $L$. Markers indicate the energy calculated
for domain patterns in Fig. \ref{fig:3}(a-d) and Fig. \ref{fig:6}(a-d)
using phase-field simulations. }}

\end{figure}

\textcolor{black}{\small{}Here $\gamma_{90}\sim6\mathrm{mJ/m^{2}}$
and $\gamma_{180}\sim13\mathrm{mJ/m^{2}}$ are the $90^{\circ}$ and
$180^{\circ}$ domain wall energies respectively. $A_{90}$ and $A_{180}$
are the corresponding domain wall areas within the periodic cell.
Dividing through by the periodic cell volume and using the normalization
$\tilde{\psi}=(\psi_{0}-\psi)/\psi_{0}$, the variation of the normalized
free energy with scale can be compared for the different patterns,
see Fig. \ref{fig:8}.}{\small \par}

\textcolor{black}{\small{}When the periodic cell size is $L<100\mathrm{nm}$,
the contributions from gradient energy become significant, increasing
the energy of each of the rank-1 and rank-2 domain patterns. Conversely,
for periodic cells a few micrometers in size, the elastic energy contribution
is dominant, and the energy density approaches a constant for the
rank-2 domain patterns with non-zero elastic energy. For the herringbone
pattern, \textquotedblleft 1234\textquotedblright{} and the rank-1
laminates, the domains are almost stress-free and so the gradient
energy is the dominant contribution over a wide range of scale: then
$\tilde{\psi}\propto L^{-1}$. The percentage of gradient energy contribution
for complex domain patterns such as the checkerboard pattern \textquotedblleft 1324\textquotedblright{}
is low at the microscale, but they have high overall energy because
of residual stresses. }{\small \par}

\textcolor{black}{\small{}The phase field calculations of free energy,
carried out at $L=40\mathrm{nm}$ and $20\mathrm{nm}$ for various
laminates are marked in Fig. \ref{fig:8}. It was not practically
feasible to carry out phase-field simulations over a wide range of
scales because of the computational cost, which increases approximately
as $L^{4}$. Reducing the periodic cell size much below $L=20\mathrm{nm}$
is also problematic because the spacing between domain walls becomes
comparable to the domain wall width \textendash{} then the domains
become unstable. This lower limit of stability is indicated by a transition
to dashed lines in Fig. \ref{fig:8}. }{\small \par}

\textcolor{black}{\small{}Since the area and volume contributions
differ from pattern to pattern, some scale dependent cross-over in
energies is expected, as seen in Fig. \ref{fig:8}. However, the energy
estimates are not sufficiently precise to quantify the length scales
at which cross-over occurs. In any case, we can see that the simpler,
stress-free rank-1 laminates have lower energy than the rank-2 patterns
for all sizes of the periodic cell. Stability of the laminates under
zero external load conditions does not correspond to a global energy
minimum, but rather a local minimum or neutral equilibrium state. }{\small \par}

\textcolor{black}{\small{}Fig. \ref{fig:8} provides a qualitative
insight into the domain pattern energies. Rank-2 domain patterns which
were unstable at the 40nm scale under zero external load conditions
reduced their gradient energy and elastic energy contributions by
evolving into simpler patterns such as rank-1 stripes, or forming
a monodomain. By contrast, the herringbone domain pattern \textquotedblleft 1234\textquotedblright{}
reached equilibrium although having substantial domain wall energy.
Closer examination reveals that this domain pattern is easily dissolved:
for example, application of external electric field $E_{2}=-0.05E_{0}\sim1\mathrm{MV/m}$
dissolves this domain pattern into a monodomain. These findings are
important because they suggest that the rank-2 patterns other than
the well-known herringbone pattern, are relatively unlikely to form
at microscale, though they are occasionally seen \cite{key-18}. This
is due to their having high energy from residual stress. However,
at the nanoscale the energies of rank-2 and rank-1 laminates become
rather similar. At this scale, the more complex patterns are still
energetically unfavourable, but there may be a possibility to stabilize
them with stress/strain conditions, or electrical conditions. This
opens an exciting possibility of domain engineering at the nanoscale
that will be explored further in section III-C below. }{\small \par}

\begin{figure}
\begin{centering}
\includegraphics[width=1\columnwidth]{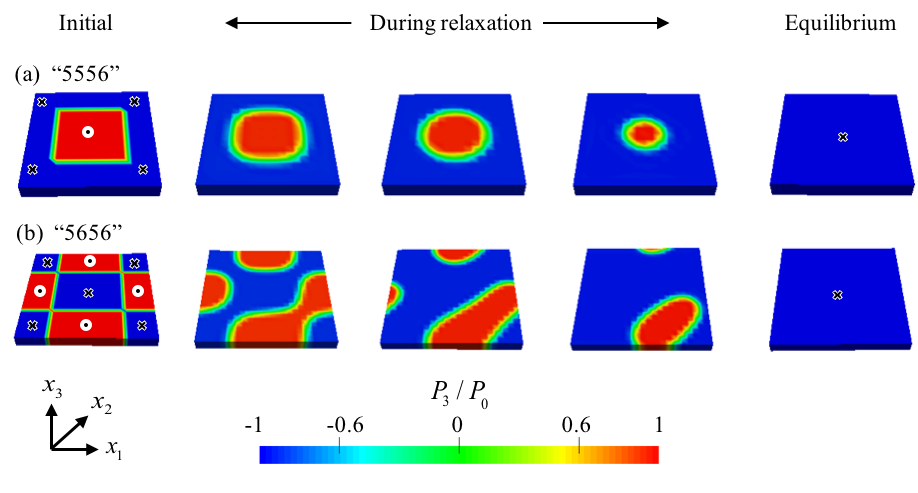}
\par\end{centering}
\caption{\textcolor{black}{\small{}\label{fig:9}Testing the stability of rank-2
domain patterns with out-of-plane polarizations $\pm P_{3}$ under
zero external load conditions: (a) laminate \textquotedblleft 5556\textquotedblright{}
and (b) laminate \textquotedblleft 5656\textquotedblright . }}

\end{figure}

\begin{figure}
\begin{centering}
\includegraphics[width=1\columnwidth]{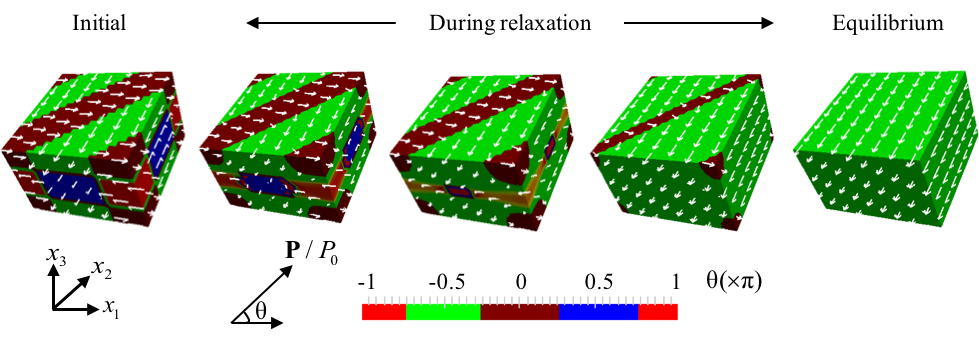}
\par\end{centering}
\textcolor{black}{\small{}\caption{\textcolor{black}{\small{}\label{fig:10}Testing the stability of
rank-2 domain pattern \textquotedblleft 1423\textquotedblright{} with
3D domain wall arrangement under zero external load conditions. }}
}{\small \par}

\end{figure}

\textcolor{black}{\small{}Next, we test the stability of three rank-2
domain patterns with out-of-plane polarizations under zero external
load conditions as described for a 3D model by Eq. \ref{eq:3}. In
each case we use the minimum periodic cell size that allows the pattern
to be represented with sufficient elements to resolve the domain walls.
Two of these domain patterns, namely \textquotedblleft 5556\textquotedblright{}
and \textquotedblleft 5656\textquotedblright , see Fig. \ref{fig:9}(a, b),
have a 2D arrangement of domain walls, but out-of-plane polarizations,
parallel to $x_{3}$. For these patterns a prismatic plate-like periodic
cell is used. The third pattern \textquotedblleft 1423\textquotedblright ,
see Fig. \ref{fig:10}, has a fully 3D domain wall arrangement, requiring
a thicker periodic cell. }{\small \par}

\textcolor{black}{\small{}The laminate \textquotedblleft 5556\textquotedblright{}
is initialized with its polarization values on a periodic cell of
dimension $16\mathrm{nm}\times16\mathrm{nm}\times2\mathrm{nm}$, see
Fig. \ref{fig:9}a. During relaxation, the $180^{\circ}$ domain walls
curve to enclose a cylindrical $P_{3}$ domain. The radius of this
cylindrical domain reduces to zero, leaving a monodomain at equilibrium.
The volume average free energy and average axial components of strain
and polarization adopted by this domain pattern in the settled state
are given in Table II. It is interesting to note that domain patterns
similar to the \textquotedblleft 5556\textquotedblright{} laminate,
though less regular, are observed experimentally \cite{key-30,key-71}.
The current phase-field model lacks lattice friction, which is likely
to be a factor in stabilising this domain pattern. }{\small \par}

\textcolor{black}{\small{}For interest, we test the ability of an
external electric field $E_{3}$ to hold a nanoscale cylindrical domain,
such as that formed during the relaxation of pattern \textquotedblleft 5556\textquotedblright ,
in equilibrium. The simulation is initialized with the \textquotedblleft 5556\textquotedblright{}
pattern but subject to the condition }{\small \par}

\textcolor{black}{\small{}
\begin{equation}
\phi(x_{1},x_{2},D)=\phi(x_{1},x_{2},0)-E_{3}D.\label{eq:6}
\end{equation}
}{\small \par}

\textcolor{black}{\small{}Other boundary conditions are as in Eq.
\ref{eq:3}, giving zero average stress. While it was possible to
hold cylindrical domains in equilibrium with electric field, the pattern
was unstable to small perturbations in applied field strength, dissolving
to a monodomain. }{\small \par}

\textcolor{black}{\small{}The periodic domain pattern \textquotedblleft 5656\textquotedblright{}
shown in Fig. \ref{fig:9}b is imposed on a periodic cell of size $24\mathrm{nm}\times24\mathrm{nm}\times2\mathrm{nm}$,
with corresponding polarization values in the initial state. This
domain pattern also dissolves into a monodomain. During relaxation,
symmetry is broken and the $180^{\circ}$ domain walls curve to form
loops, each enclosing a $P_{3}$ domain. This reduces the domain wall
area as discussed by Tagantsev }\textit{\textcolor{black}{\small{}et al.}}\textcolor{black}{\small{}
\cite{key-27} The eventual collapse to a monodomain is similar to
that of the \textquotedblleft 5556\textquotedblright{} pattern. Patterns
with intersecting $180^{\circ}$ domain walls, like \textquotedblleft 5656\textquotedblright{}
are rarely observed. }{\small \par}

\textcolor{black}{\small{}}
{\small \par}

\textcolor{black}{\small{}}
\begin{table}
\noindent %
\begin{tabular*}{1\columnwidth}{@{\extracolsep{\fill}}>{\raggedright}p{0.15\columnwidth}>{\centering}p{0.2\columnwidth}cccc}
\toprule 
{\scriptsize{}Rank \textendash{} 2 periodic laminates} &
{\scriptsize{}Average free energy } &
\multicolumn{2}{>{\centering}p{0.2\columnwidth}}{{\scriptsize{}Average strain components $\tilde{\epsilon}_{ij}=\left\langle \epsilon_{ij}\right\rangle /\epsilon_{0}$}} &
\multicolumn{2}{>{\centering}p{0.28\columnwidth}}{{\scriptsize{}Average polarization components $\tilde{P}_{i}=\left\langle P_{i}\right\rangle /P_{0}$}}\tabularnewline
 &
{\scriptsize{}$\tilde{\psi}$} &
{\scriptsize{}$\tilde{\epsilon}_{11}=\tilde{\epsilon}_{22}$} &
{\scriptsize{}$\tilde{\epsilon}_{33}$} &
{\scriptsize{}$\tilde{P}_{1}=\tilde{P}_{2}$} &
{\scriptsize{}$\tilde{P}_{3}$}\tabularnewline
\midrule
{\scriptsize{}(a) ``5556''} &
{\scriptsize{}0.39} &
{\scriptsize{}$-0.26$} &
{\scriptsize{}0.82} &
{\scriptsize{}0.00} &
{\scriptsize{}$-0.50$}\tabularnewline
{\scriptsize{}(b) ``5656''} &
{\scriptsize{}0.70} &
{\scriptsize{}$-0.27$} &
{\scriptsize{}0.86} &
{\scriptsize{}0.00} &
{\scriptsize{}0.00}\tabularnewline
{\scriptsize{}(c) ``1423''} &
{\scriptsize{}0.74} &
{\scriptsize{}0.24} &
{\scriptsize{}$-0.19$} &
{\scriptsize{}0.00} &
{\scriptsize{}0.00}\tabularnewline
\bottomrule
\end{tabular*}

\textcolor{black}{\small{}\caption{\textcolor{black}{\small{}\label{tab:2}The normalized volume average
free energy $\tilde{\psi}$, average axial strains $\tilde{\epsilon}_{ij}$,
and polarization $\tilde{P}_{i}$ adopted by rank-2 periodic domain
patterns with out-of-plane polarization in their settled states. }}
}{\small \par}
\end{table}
{\small \par}

\textcolor{black}{\small{}The \textquotedblleft 1423\textquotedblright{}
domain pattern comprises layers of rank-1 stripe domain pattern arranged
in a three dimensional configuration as shown in Fig. \ref{fig:10}.
This domain pattern is initialized on a periodic cell of size $20\mathrm{nm}\times20\mathrm{nm}\times13\mathrm{nm}$.
At this periodic cell size, the gradient energy contribution is substantial,
see Table II; this destabilizes the domain pattern. However, the layered
stripe domain pattern is disclination free. It has been observed in
experiments \cite{key-18,key-27}, and we expect that this domain
pattern could reach equilibrium at a larger size of the periodic cell,
or with suitable boundary constraints. In other work, the pattern
was found stable with a 33nm cell size, whose boundaries were constrained
to match a fixed strain field \cite{key-79}.}{\small \par}

\textcolor{black}{\small{}A fourth rank-2 laminate with out-of-plane
polarization, \textquotedblleft 1325\textquotedblright{} contains
domains polarized along all three coordinate axes \cite{key-18},
and requires a minimum periodic cell size of about 60nm. This pattern
is not tested here, due to computational limitations, but is the subject
of ongoing work. }{\small \par}

\textcolor{black}{\small{}To conclude the discussion of rank-2 laminates
with zero external load conditions, our key finding is that the majority
of these complex laminates are not stable at the nanoscale. The exception
is the well-known herringbone domain pattern. The main reasons for
the instability of most of the patterns appear to be the high contribution
to the overall energy from either disclinations or domain walls. Nevertheless,
these structures, if they could be stabilized, offer the possibility
of nano-engineering of domain configurations. We thus explore next
the possibility of stabilizing these patterns using external loading. }{\small \par}

\subsection*{{\footnotesize{}Stabilization by external loading}}

\textcolor{black}{\small{}Are there loading conditions that can stabilize
the rank-2 patterns? To explore this question, we start by considering
the average of the spontaneous strain in each of the three laminates
\textquotedblleft 1323\textquotedblright , \textquotedblleft 1434\textquotedblright{}
and \textquotedblleft 1324\textquotedblright , found unstable under
zero external load conditions. Suppose a given pattern contains $n$
domains, numbered $k=1\ldots n$, each having a volume fraction $f_{k}$
and normalized spontaneous strain tensor $\epsilon_{ij}^{k}$. Neglecting
domain wall volume, the volume average of spontaneous strain in the
periodic cell is}{\small \par}

\textcolor{black}{\small{}
\[
\overline{\epsilon}_{ij}=\sum_{k=1}^{n}f_{k}\epsilon_{ij}^{k}\qquad(0\leq f_{k}\leq1,\sum f_{k}=1)
\]
}{\small \par}

\textcolor{black}{\small{}In the simulations of section III-B, the
periodic cell under zero load conditions given by Eq. \ref{eq:3},
was allowed to adopt any value of average strain in order to minimize
energy. The resulting equilibrium states have different domain volume
fractions from the initial states and}{\small \par}

\textcolor{black}{\small{}}
\begin{figure}
\begin{centering}
\textcolor{black}{\small{}\includegraphics[width=1\columnwidth]{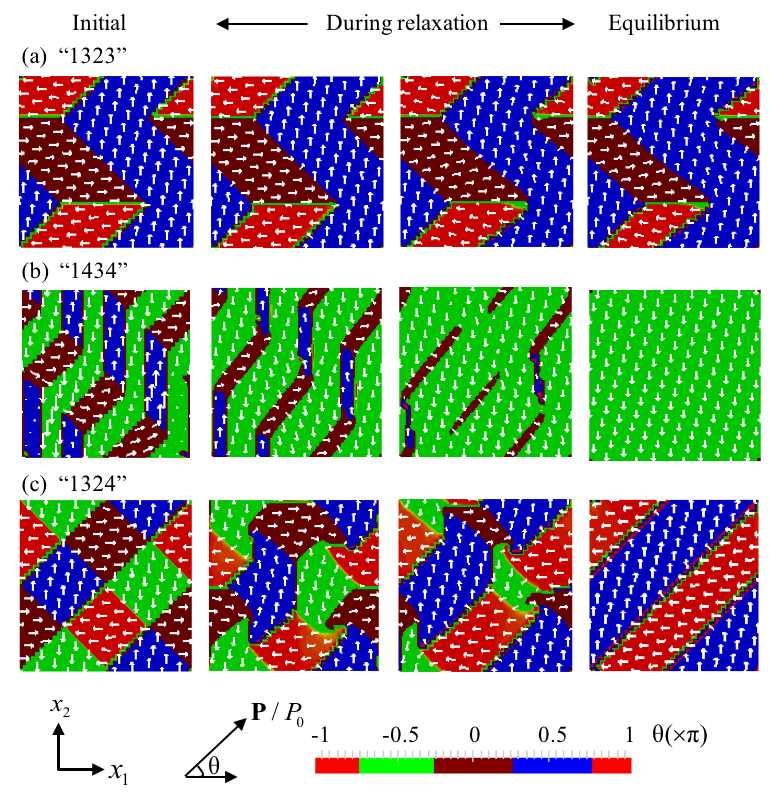}}
\par\end{centering}{\small \par}
\textcolor{black}{\small{}\caption{\textcolor{black}{\small{}\label{fig:11}Evolution of rank-2 laminates
(a) \textquotedblleft 1323\textquotedblright{} (b) \textquotedblleft 1434\textquotedblright{}
(c) \textquotedblleft 1324\textquotedblright , under fixed average
strain boundary conditions.}}
}{\small \par}

\end{figure}
{\small \par}

\noindent \textcolor{black}{\small{}hence the average spontaneous
strain changes during relaxation. We next conduct simulations in which
the average strain is fixed at the $\bar{\epsilon}_{ij}$ value of
the initial state, using the boundary conditions }{\small \par}

\textcolor{black}{\small{}
\begin{equation}
\begin{array}{c}
u_{1}(L,x_{2})\mid_{R}=u_{1}(0,x_{2})\mid_{R_{m}}+\overline{\epsilon}_{11}L,\\
u_{2}(x_{1},L)\mid_{S}=u_{2}(x_{1},0)\mid_{S_{m}}+\overline{\epsilon}_{22}L,
\end{array}\label{eq:8}
\end{equation}
}{\small \par}

\noindent \textcolor{black}{\small{}in place of the corresponding
displacement boundary conditions in Eq. \ref{eq:3}. All other boundary
conditions remain unchanged. As before, plane simulations were conducted,
enforcing $P_{3}=0$ and fixing $\epsilon_{33}=\epsilon_{0}^{t}$.
The resulting pattern evolution is shown in Fig. \ref{fig:11}: fixing
the average strain stabilizes the \textquotedblleft 1323\textquotedblright{}
laminate, while \textquotedblleft 1434\textquotedblright{} and \textquotedblleft 1324\textquotedblright{}
still collapse into lower energy patterns. In the case of the \textquotedblleft 1324\textquotedblright{}
pattern, the $90^{\circ}$ stripe pattern that forms matches the imposed
$\bar{\epsilon}_{ij}$ without stress, while the \textquotedblleft 1434\textquotedblright{}
collapses to a stressed monodomain. }{\small \par}

\textcolor{black}{\small{}A further attempt to stabilize the \textquotedblleft 1434\textquotedblright{}
and \textquotedblleft 1324\textquotedblright{} patterns can be made
by imposing a fixed average polarization state to match the average
spontaneous polarization of the initial state. This discourages the
formation of monodomain or simple stripe patterns because they do
not match the initial average polarization. Defining $P_{i}^{k}$
as the normalized spontaneous polarization of the $k$th domain and
neglecting domain wall volume, the volume average polarization in
the periodic cell is }{\small \par}

\textcolor{black}{\small{}
\begin{equation}
\overline{P}_{i}=\sum_{k=1}^{n}f_{k}P_{i}^{k}\qquad(0\leq f_{k}\leq1,\sum f_{k}=1)\label{eq:9}
\end{equation}
}{\small \par}

\textcolor{black}{\small{}This is applied to the periodic cell by
replacing the zero electric field condition in Eq. \ref{eq:3} with }{\small \par}

\textcolor{black}{\small{}
\begin{equation}
\phi(L,x_{2})\mid_{R}-\phi(L,0)\mid_{B}=\phi(0,x_{2})\mid_{R_{m}}-\phi(0,0)\mid_{A}\label{eq:10}
\end{equation}
}{\small \par}

\begin{figure}
\begin{centering}
\includegraphics[width=1\columnwidth]{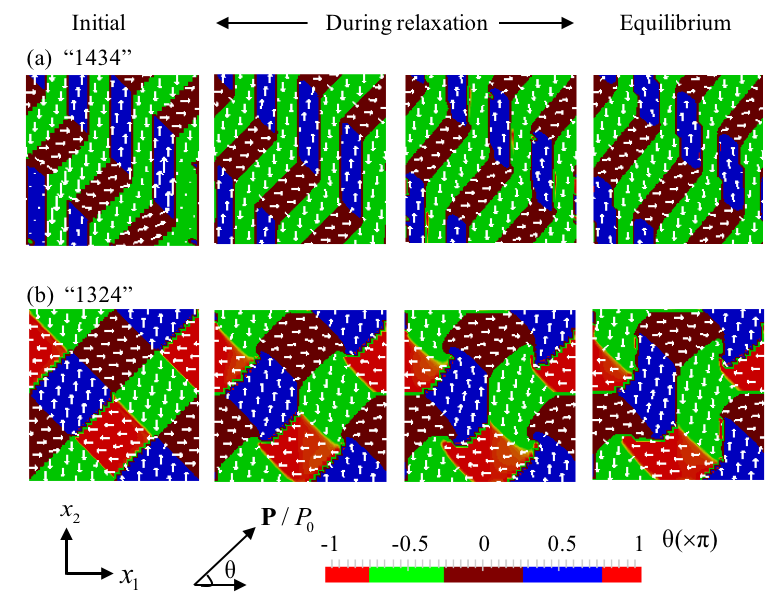}
\par\end{centering}
\textcolor{black}{\small{}\caption{\textcolor{black}{\small{}\label{fig:12}Evolution of the rank-2 domain
patterns (a) \textquotedblleft 1434\textquotedblright{} (b) \textquotedblleft 1324\textquotedblright ,
under fixed average polarization condition. }}
}{\small \par}

\end{figure}

\noindent \textcolor{black}{\small{}and similar conditions for typical
node $S$. These boundary conditions allow an arbitrary, but periodic,
electric field. A uniform charge density $q=-\bar{P}_{i}n_{i}$ (where
$n_{i}$ is the outward surface normal) can then be added on all boundary
nodes to enforce the average polarization. The other boundary conditions
are as given in Eq. \ref{eq:3}. Note that in the checkerboard pattern
\textquotedblleft 1324\textquotedblright , $\bar{P}_{i}=0$. The resulting
simulations are shown in Fig. \ref{fig:12}(a-b), where it can be
seen that the imposition of an average polarization does stabilize
pattern \textquotedblleft 1434\textquotedblright . However, the checkerboard
pattern \textquotedblleft 1324\textquotedblright{} undergoes some
pattern change before stabilizing. During}{\small \par}

\textcolor{black}{\small{}}
{\small \par}

\noindent \textcolor{black}{\small{}relaxation, $180^{\circ}$ domain
walls appear at the $90^{\circ}$ domain junctions, forming a \textquoteleft displaced\textquoteright{}
checkerboard pattern. Similar features at domain junctions have been
imaged in BaTiO$_{3}$ lamellae by McGilly }\textit{\textcolor{black}{\small{}et
al.}}\textcolor{black}{\small{} \cite{key-15} The simulations with
imposed average strain or polarization demonstrate that periodic laminations
of domains can be stabilized by external loading: if domains form
in a region of crystal that is mechanically constrained, or has charged
surfaces, nanoscale patterns such as these could form. Thus in a prestressed
thin film or particle it may be possible to engineer specific nanoscale
patterns through the control of the stress/strain and external charge
state. This suggests an exciting route towards generating specific
arrangements of domains with useful functional properties. However
it should be noted that the existence of a stable pattern does not
imply that the material will adopt that pattern from arbitrary starting
conditions. In the simulations so far, an idealized pattern was imposed
as a starting state: this would not normally be possible in a practical
device. This leads us to explore whether complex patterns can be generated
starting from known simple patterns, and whether they can be transformed
or switched by applied fields. }{\small \par}

\subsection*{{\footnotesize{}Switching and transformation of patterns}}

\begin{figure}
\begin{centering}
\includegraphics[width=1\columnwidth]{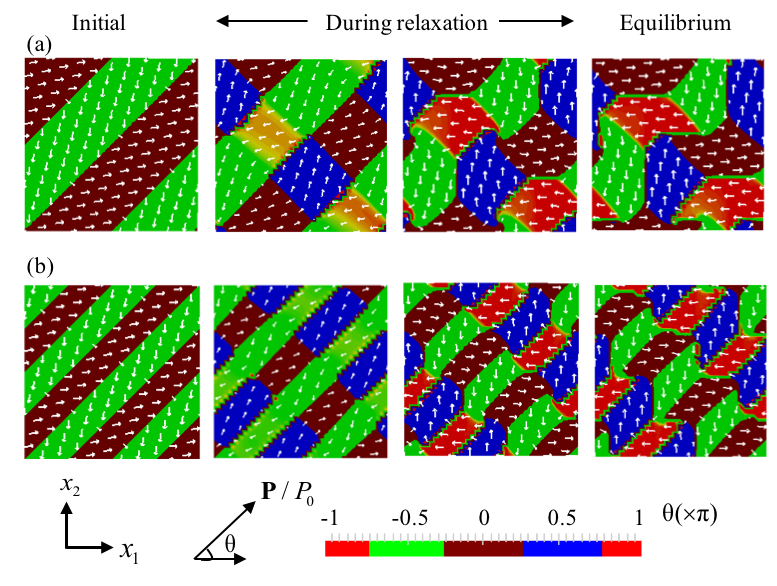}
\par\end{centering}
\textcolor{black}{\small{}\caption{\textcolor{black}{\small{}\label{fig:13}Evolution of the rank-1 $90^{\circ}$
stripe domain pattern under zero average polarization (a) $s=14.1\mathrm{nm}$
(b) $s=7.1\mathrm{nm}$.}}
}{\small \par}

\end{figure}

\textcolor{black}{\small{}Can complicated patterns, such as the rank-2
laminates, develop under some conditions, from simpler patterns, such
as the rank-1 stripes? To explore this, we study the effect of imposing
zero average polarization on rank-1 laminates \textquotedblleft 14\textquotedblright{}
with $s=14.1\mathrm{nm}$ and $s=7.1\mathrm{nm}$, as shown in Fig.
\ref{fig:13}. The simulations are initialised with the $90^{\circ}$
stripe domain patterns at equilibrium from Fig. \ref{fig:3}(c-d).
Note that these stripe patterns possess a net polarization field and
only $90^{\circ}$ domain walls. If we now impose $\bar{P}_{i}=0$,
some $180^{\circ}$ domains are expected to form to accommodate the
average polarization. In the simulations, during relaxation, the polarization
in the domains reduces to near zero magnitude and new domains nucleate
such that rank-2 laminate patterns begin to evolve. At first, the
strong electric fields generated by imposing a zero average polarization
state cause the local polarization vectors to bend far away from the
crystallographic axes \textendash{} a high energy vortex-like state
that must relax further. But as the system approaches equilibrium
the polarization in each new domain returns close to the spontaneous
value and aligns with the crystal axes. For $s=14.1\mathrm{nm}$,
see Fig. \ref{fig:13}a a \textquoteleft displaced\textquoteright{}
checkerboard pattern as seen in Fig. \ref{fig:12}b is found at equilibrium. }{\small \par}

\textcolor{black}{\small{}}
{\small \par}

\textcolor{black}{\small{}The periodic pattern in Fig. \ref{fig:13}a
is similar to the experimental observation of a periodic array of
flux-closure domains in PbTiO$_{3}$ reported by Tang }\textit{\textcolor{black}{\small{}et
al.}}\textcolor{black}{\small{} \cite{key-28} While with $s=7.1\mathrm{nm}$,
see Fig. \ref{fig:13}b a herringbone domain pattern is stabilised.
Care is needed in interpreting these results: observe that scaling
of the cell size in these simulations (or equivalently, changing the
domain spacing) leads to different equilibrium states. This happens
because the imposed conditions have forced the polarization, at first,
far from the spontaneous states making the subsequent evolution of
domains highly sensitive to the kinetics, here represented crudely
through the polarization viscosity. Thus the path of domain evolution
during switching is to some extent an artefact of the model. Also,
from an energetic point of view, the nucleation of the new domains
is less costly in the simulation if the pattern adopts the minimum
number of periods that will fit into the simulation cell. This explains
the way the individual domains form new layers with the minimum possible
spatial frequency. This is an artefact of the cell size: in nature
there is no imposed cell and the new spacing must arise purely from
a balance between the gradient and bulk energies. Nevertheless, the
simulations in Fig. \ref{fig:13} indicate the opportunity to engineer
complex patterns such as the rank-2 laminates by starting with a simple
rank-1 pattern and applying suitable external conditions that force
the pattern to nucleate new domains. The length-scale of the initial
pattern then controls or guides the generation of the new pattern.
We have given only a few examples to illustrate the process, but expect
that similar methods could enable the generation of a range of patterns. }{\small \par}

\textcolor{black}{\small{}Finally, we give an example of the effect
of electric field on a strain stabilized laminate. Once again, switching
is involved, so the cell size and kinetics could affect the }{\small \par}

\textcolor{black}{\small{}}
\begin{figure}
\begin{centering}
\textcolor{black}{\small{}\includegraphics[width=1\columnwidth]{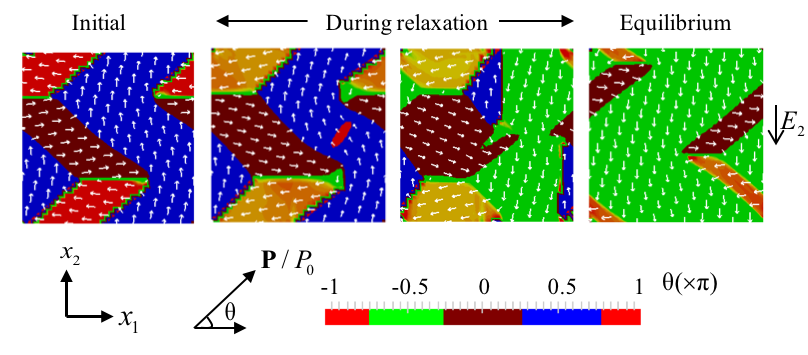}}
\par\end{centering}{\small \par}
\textcolor{black}{\small{}\caption{\textcolor{black}{\small{}\label{fig:14}Evolution of rank-2 domain
pattern \textquotedblleft 1323\textquotedblright{} stabilized by average
strain and under electric field, $E_{2}=-20\mathrm{MV/m}$. }}
}{\small \par}

\end{figure}
{\small \par}

\noindent \textcolor{black}{\small{}simulations. Taking as initial
state the stabilized \textquotedblleft 1323\textquotedblright{} laminate
from Fig. \ref{fig:11}a, additional loading in the form of an electric
field of magnitude $0.9E_{0}=20\mathrm{MV/m}$ is applied, and the
pattern is relaxed towards a new equilibrium state. If the field is
applied along the $\pm x_{1}$ direction the pattern collapses into
$90^{\circ}$ stripe domains, by growing whichever domain is aligned
to the applied electric field. Electric field in the $+x_{2}$ direction
expands the domain polarized in this direction, but the pattern is
otherwise unchanged. However, Fig. \ref{fig:14} shows the interesting
case where electric field is applied along the $-x_{2}$ direction
and there is no domain in the initial state that is aligned with the
applied electric field. The pattern undergoes a transition which produces
laminate \textquotedblleft 1424\textquotedblright , a symmetry related
pattern in the same family. From symmetry, the process can be reversed
by subsequent application of electric field in the $+x_{2}$ direction,
suggesting a mechanism for cyclic polarization switching of the nanoscale
periodic patterns. Such switchable nanoscale patterns could have great
potential for memory elements, or tunable devices. }{\small \par}

\section*{{\small{}Conclusion}}

\textcolor{black}{\small{}The stability of rank-1 and rank-2 laminates
comprising of nanoscale domains formed in tetragonal ferroelectrics
was explored using a phase-field model, taking account of energy contributions
from gradient energy and elastic strain energy. With zero external
load conditions, the rank-1 laminates (simple stripe patterns) were
found to be in neutral equilibrium, while the more complex rank-2
laminates were on the whole unstable, with the exception of the herringbone
domain pattern. Rank-2 nanoscale domain patterns with multiple domain
walls and disclinations were found to possess high energy density,
which destabilized them. The effect of scaling the rank-1 and rank-2
laminates was qualitatively discussed. It was found that the unstable
rank-2 domain patterns could be equilibrated under external loads
such as electric field, average strain or polarization, representative
residual states of stresses and electric fields that are commonly
present or can be imposed on ferroelectric crystals. Finally, the
simulations indicated possible routes to generate rank-2 laminated
domain patterns from simpler rank-1 domain patterns, and how polarization
reversal could occur in nanoscale periodic patterns. }{\small \par}

\section*{{\small{}Acknowledgements}}

\textcolor{black}{\small{}The authors would like to acknowledge the
use of the University of Oxford Advanced Research Computing (ARC)
facility in carrying out this work. Ananya Renuka Balakrishna acknowledges
the support of a scholarship from the Felix Trust. The authors also
wish to thank Prof. C. M. Landis for help in providing program codes
and advice. }{\small \par}

\end{document}